\documentclass[3p,times,twocolumn]{elsarticle}

\usepackage{ecrc}

\volume{00}

\firstpage{1}

\journalname{Nuclear Physics B Proceedings Supplement}

\runauth{S. Ghosh}

\jid{nuphbp}

\jnltitlelogo{Nuclear Physics B Proceedings Supplement}

\usepackage{amssymb}

\usepackage[figuresright]{rotating}

\begin{document}

\begin{frontmatter}



\title{Performance of MET reconstruction and
pileup mitigation techniques in CMS }
\author{Saranya Samik Ghosh\corref{cor1}}
\ead{saranya.ghosh@cern.ch}
\cortext[cor1]{On behalf of the CMS Collaboration}
\address{TIFR, Homi Bhabha Road, Mumbai-5, INDIA}

\dochead{}

\begin{abstract}

The performance of missing transverse momentum reconstruction algorithms is presented using 8 TeV pp collision data collected with the CMS detector, corresponding to an integrated luminosity of $12.2 \pm 0.5$  $\mathrm{fb^{-1}}$. The scale and resolution of missing transverse momentum and the effects of multiple proton-proton interactions (pileup), are measured using events with an identified Z boson or isolated photon. They are in general well described by the simulation. Advanced missing transverse momentum reconstruction algorithms are also developed specifically to mitigate the effects of large numbers of pileup interactions on missing transverse momentum resolution. Using these advanced algorithms, the dependence of the missing transverse momentum resolution on pileup interactions is reduced significantly.

\end{abstract}




\end{frontmatter}


\section{Introduction}
\label{sec1Intro}

Most stable particles produced in the proton-proton collisions at the Large Hadron Collider (LHC) can be detected by the Compact Muon Solenoid (CMS) detector \cite{bib1CMS}, with neutrinos being the notable exception along with other hypothetical weakly interacting neutral particles. The presence of such weakly interacting neutral particles can be inferred from the momentum imbalance in the plane perpendicular to the beam direction, known as the missing transverse momentum (MET), denoted as $\vec{E\!\!\!/{_T}}$, with a magnitude $E\!\!\!/{_T}$.
\\
The precise measurement of the missing transverse momentum \cite{bib2pasjme12002} is very important for several physics analyses at the LHC, such as for certain measurements of standard model physics involving W bosons and the top quark and for studies contributing to the discovery of a new boson at a mass of around 125 GeV, particularly in studies dealing with the WW and $\tau\tau$ final states \cite{bib3CMShiggs}. $\vec{E\!\!\!/{_T}}$ is also a key observable for searches for physics beyond the standard model, such as supersymmetry, extra dimensions and other dark matter candidates.
\\
The reconstruction of the missing transverse momentum is highly sensitive to various detector and reconstruction effects such as detector malfunctions, miscalibration, particle momentum mismeasurement and particle misidentification. $\vec{E\!\!\!/{_T}}$ is also sensitive to the effects of multiple proton-proton interactions (pileup). In this proceeding, the performance of $\vec{E\!\!\!/{_T}}$ reconstruction is presented along with the performance of two new $\vec{E\!\!\!/{_T}}$ reconstruction algorithms that are specifically developed to reduce the effect of pileup interactions of $\vec{E\!\!\!/{_T}}$.

\section{Reconstruction of $\vec{E\!\!\!/{_T}}$ in CMS}
\label{sec2METReconstruction}

$\vec{E\!\!\!/{_T}}$ is defined as the imbalance in the transverse momentum of all the particles detected by the CMS detector. The most widely used $\vec{E\!\!\!/{_T}}$ algorithm in CMS is based on the Particle Flow (PF) algorithm \cite{bib4CMSPF}. The PF algorithm consists of reconstructing and identifying each particle
with an optimized combination of all sub-detector information. The PF $\vec{E\!\!\!/{_T}}$ is the negative of the vectorial sum over the transverse momenta of all PF particles.
\\
Three types of corrections are applied to PF $\vec{E\!\!\!/{_T}}$. Jet energy scale corrections are propagated to $\vec{E\!\!\!/{_T}}$. The difference in detector response for charged and neutral particles originating from pileup is corrected for. There are also corrections for the asymmetry in the $\phi$ distribution of $\vec{E\!\!\!/{_T}}$ that is there due to detector effects.

\section{Performance of $\vec{E\!\!\!/{_T}}$ reconstruction}
\label{sec3METPerformance}

\begin{figure}[htbp]
\centering
\includegraphics[scale=0.4]{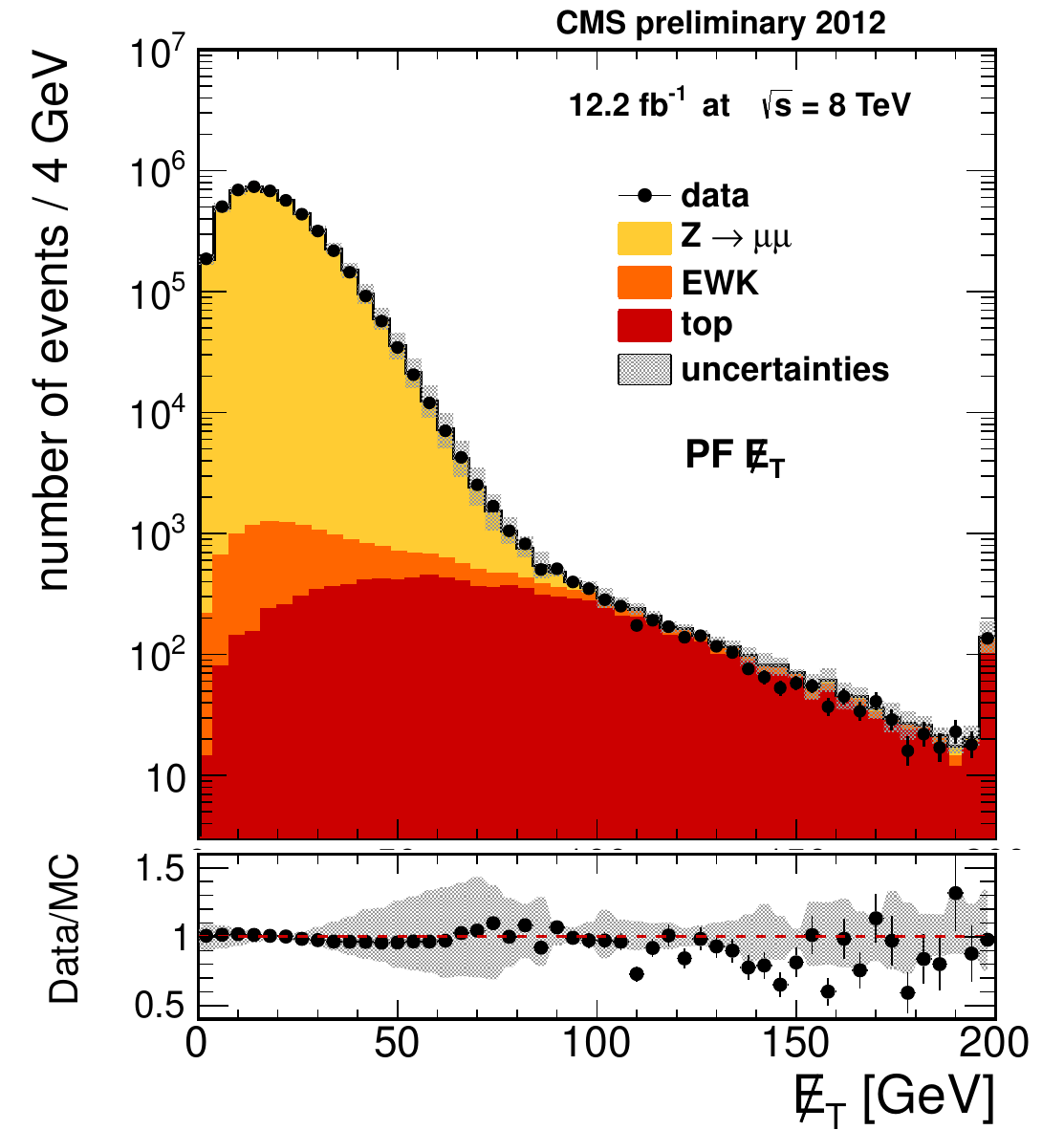}
\caption{PF $E\!\!\!/{_T}$ distribution for $Z\rightarrow\mu^{+}\mu^{-}$ candidate events in data and MC simulation.}
\label{fig:fig_PFMET_DataMC_Zmumu}
\end{figure}

Studies of performance of $\vec{E\!\!\!/{_T}}$ are based on events containing photon candidates or Z boson candidates, with the Z decaying into two electron or two muon channel. The PF $\vec{E\!\!\!/{_T}}$ distribution for $Z\rightarrow\mu^{+}\mu^{-}$ events is shown in Figure \ref{fig:fig_PFMET_DataMC_Zmumu} and shows a good agreement between data and Monte Carlo simulation.

\begin{figure}[htbp]
\centering
\includegraphics[scale=.15]{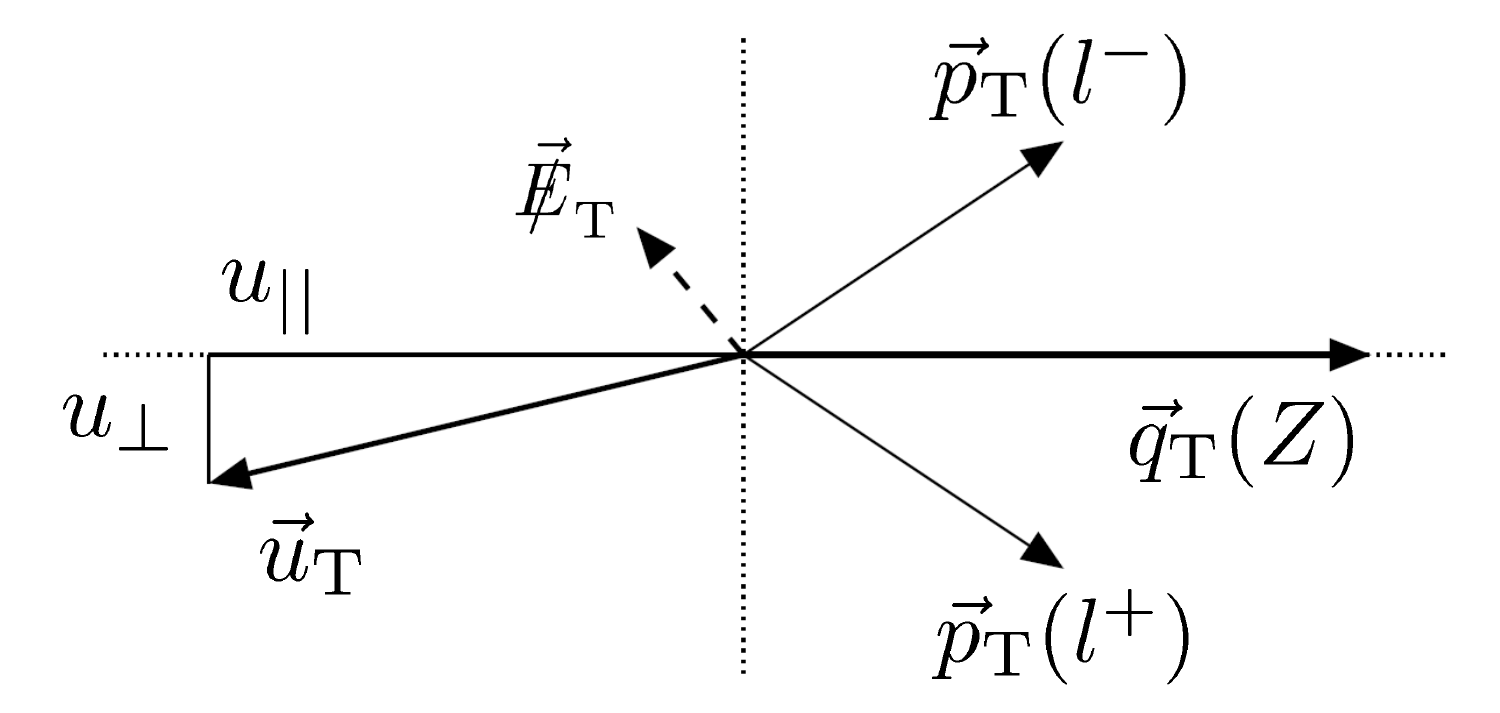}
\includegraphics[scale=.15]{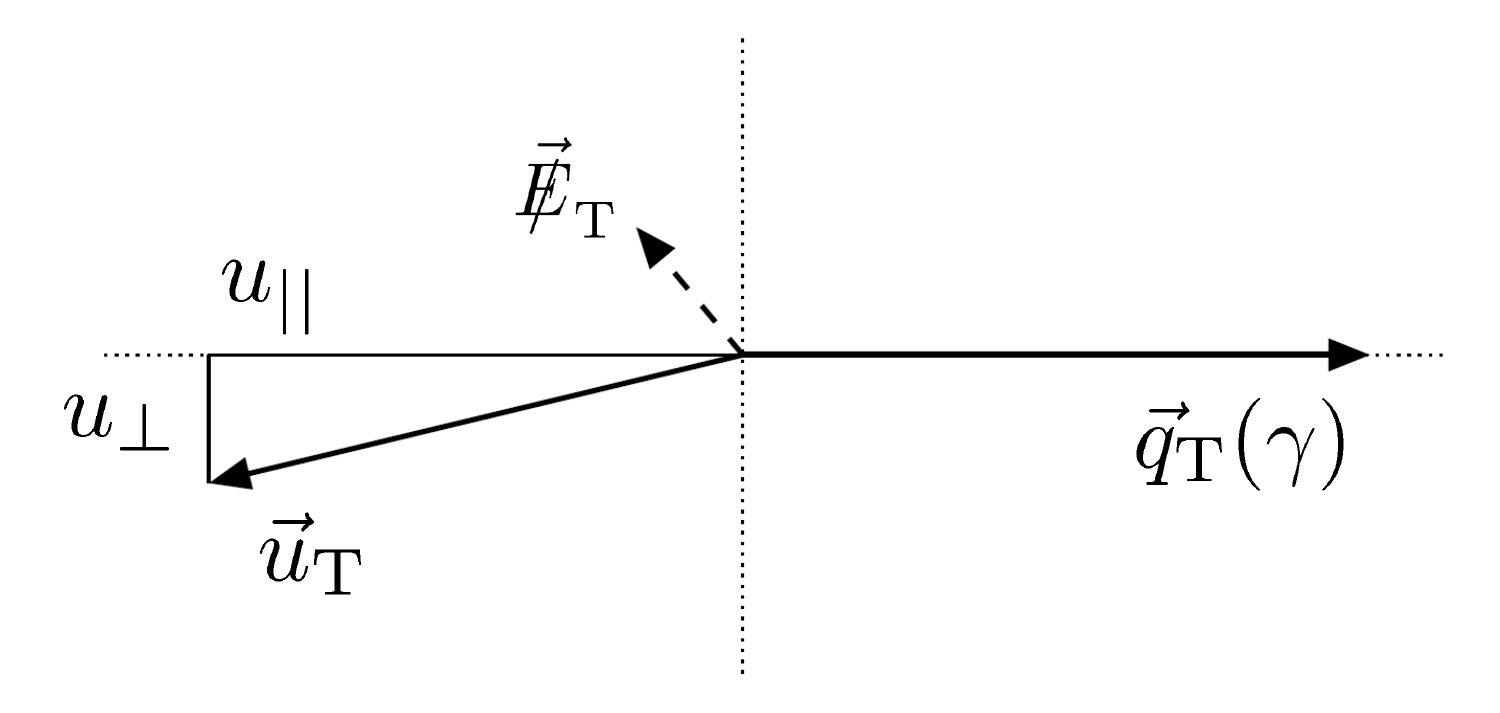}
\caption{Illustration of event kinematics in the transverse plane for $Z\rightarrow l^{+}l^{-}$ (left) and  $\gamma$ (right) events. The recoil vector $\vec{u_{T}}$ denotes the vectorial sum of all particles reconstructed in the event except for the two leptons from the Z decay (left) or the photon (right).
}
\label{fig:fig_recoilDef}
\end{figure}

\begin{figure}[htbp]
\centering
\includegraphics[scale=.2]{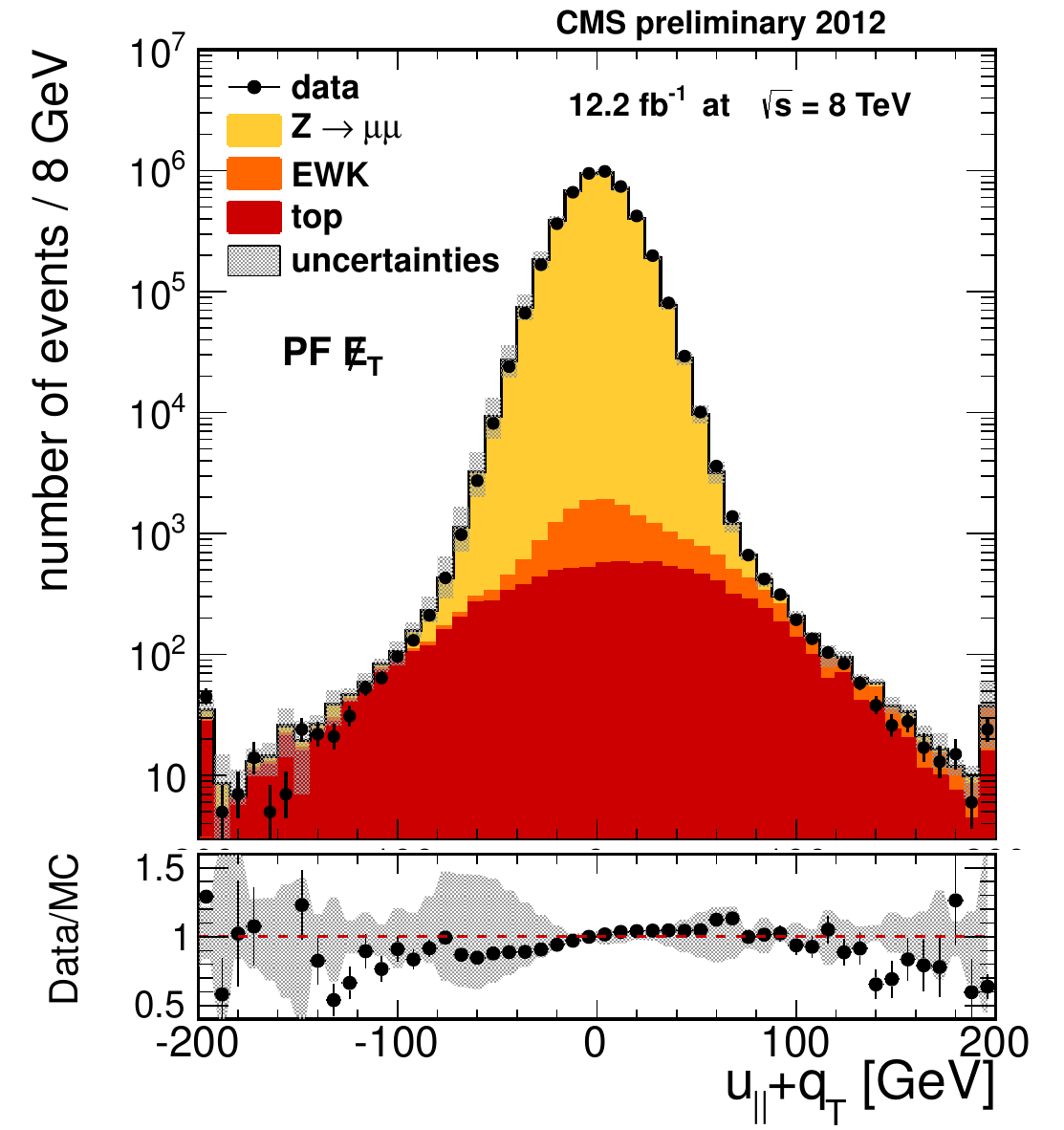}
\includegraphics[scale=.2]{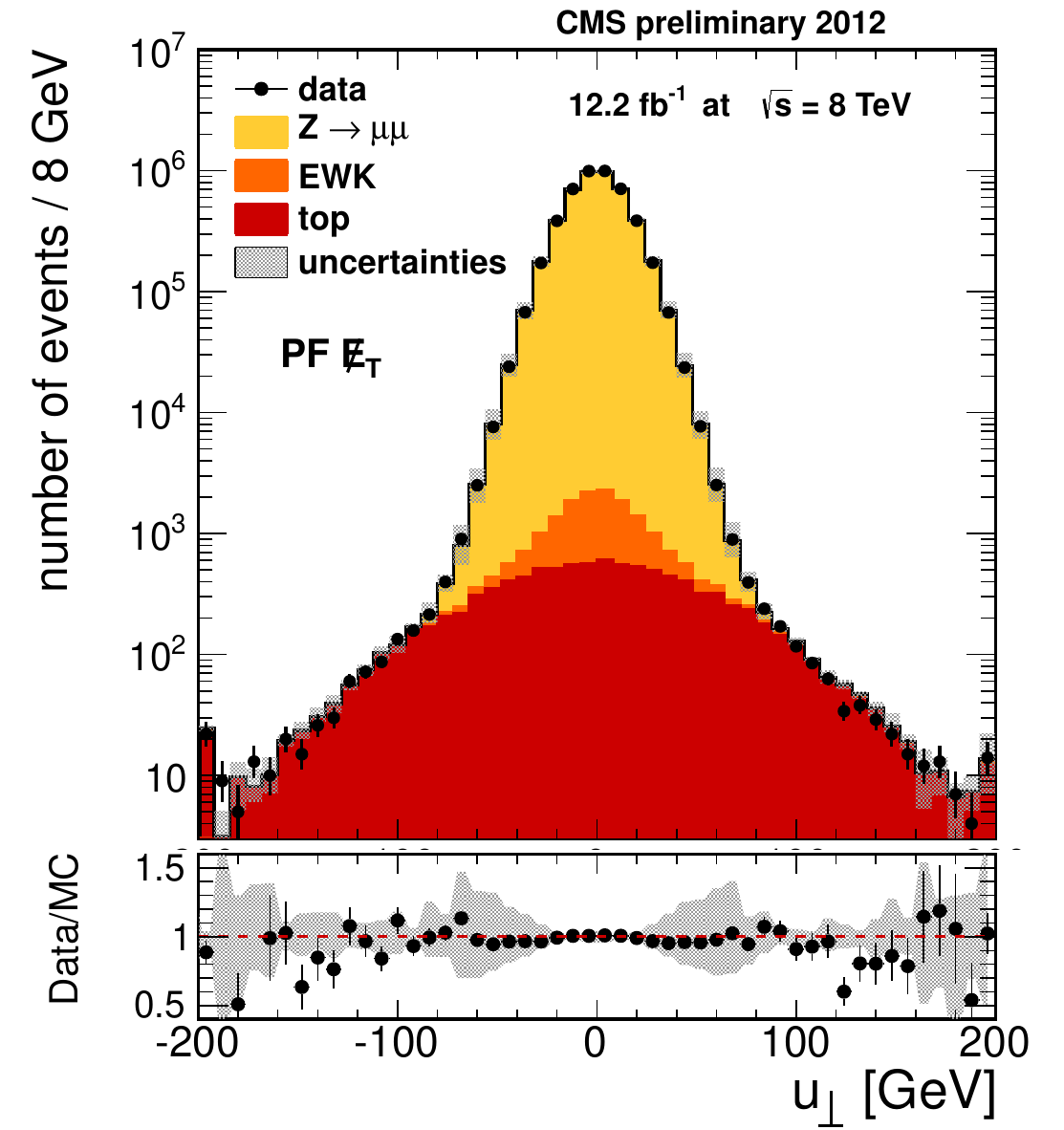}
\caption{$u_{\parallel} + q_{T}$ distribution (left) and $u_{\perp}$ distribution (right) for PF $\vec{E\!\!\!/{_T}}$ for $Z\rightarrow\mu^{+}\mu^{-}$ candidate events.
}
\label{fig:fig_Zuu_PFT1SPhiUpara}
\end{figure}

The hadronic recoil $\vec{u_{T}}$ is defined as $\vec{E\!\!\!/{_T}} + \vec{u_{T}} + \vec{q_{T}} = 0$, where  $\vec{q_{T}}$ is the photon or Z transverse momentum. The hadronic recoil projections along the $\vec{q_{T}}$ axis, $u_{\parallel}$, and along a perpendicular axis to $\vec{q_{T}}$, $u_{\perp}$, are used to estimate the performance of MET scale and resolution (Figure \ref{fig:fig_recoilDef}). The distributions of $u_{\parallel} + q_{T}$  and $u_{\perp}$  is shown in Figure \ref{fig:fig_Zuu_PFT1SPhiUpara}.

\begin{figure}[htbp]
\centering
\includegraphics[scale=0.3]
{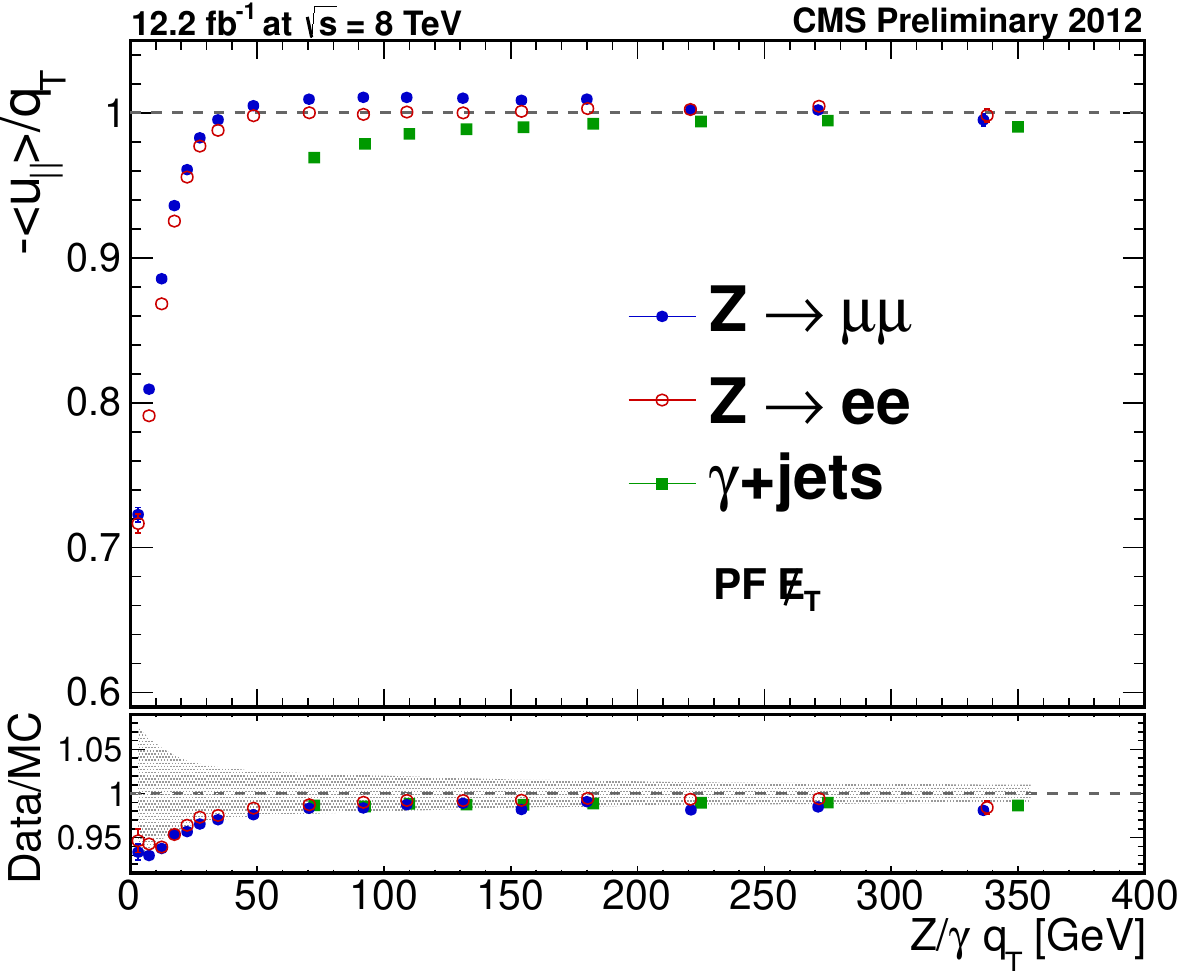}
\caption{Response for PF $\vec{E\!\!\!/{_T}}$ vs $q_{T}$ for $Z\rightarrow\mu^{+}\mu^{-}$, $Z\rightarrow e^{+}e^{-}$  and $\gamma$ events in data (on top), with the ratio of data to simulation (shown below).
}
\label{fig:fig_PFresponse}
\end{figure}

\begin{figure}[htbp]
\centering
\includegraphics[scale=.2]{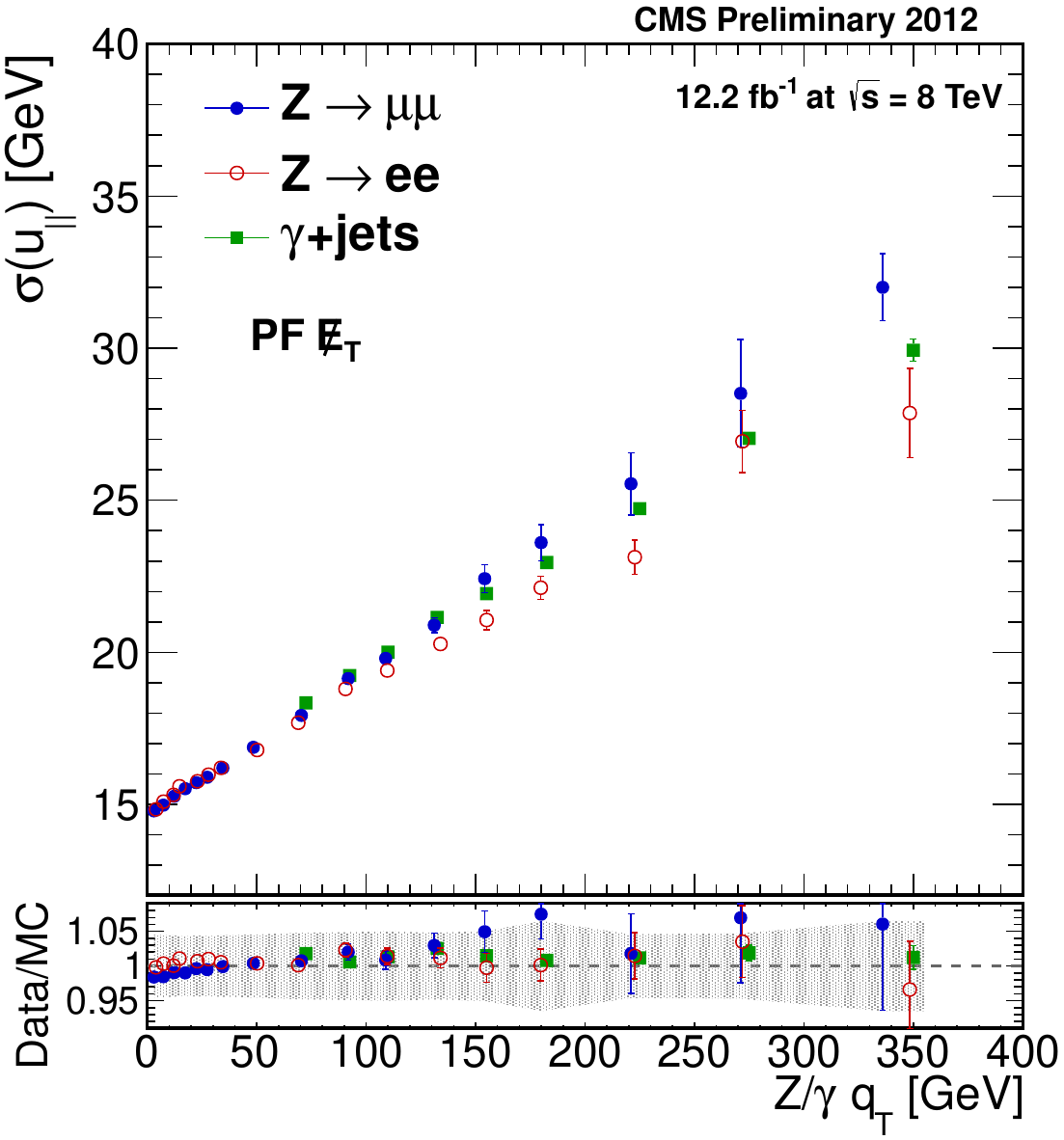}
\includegraphics[scale=.2]{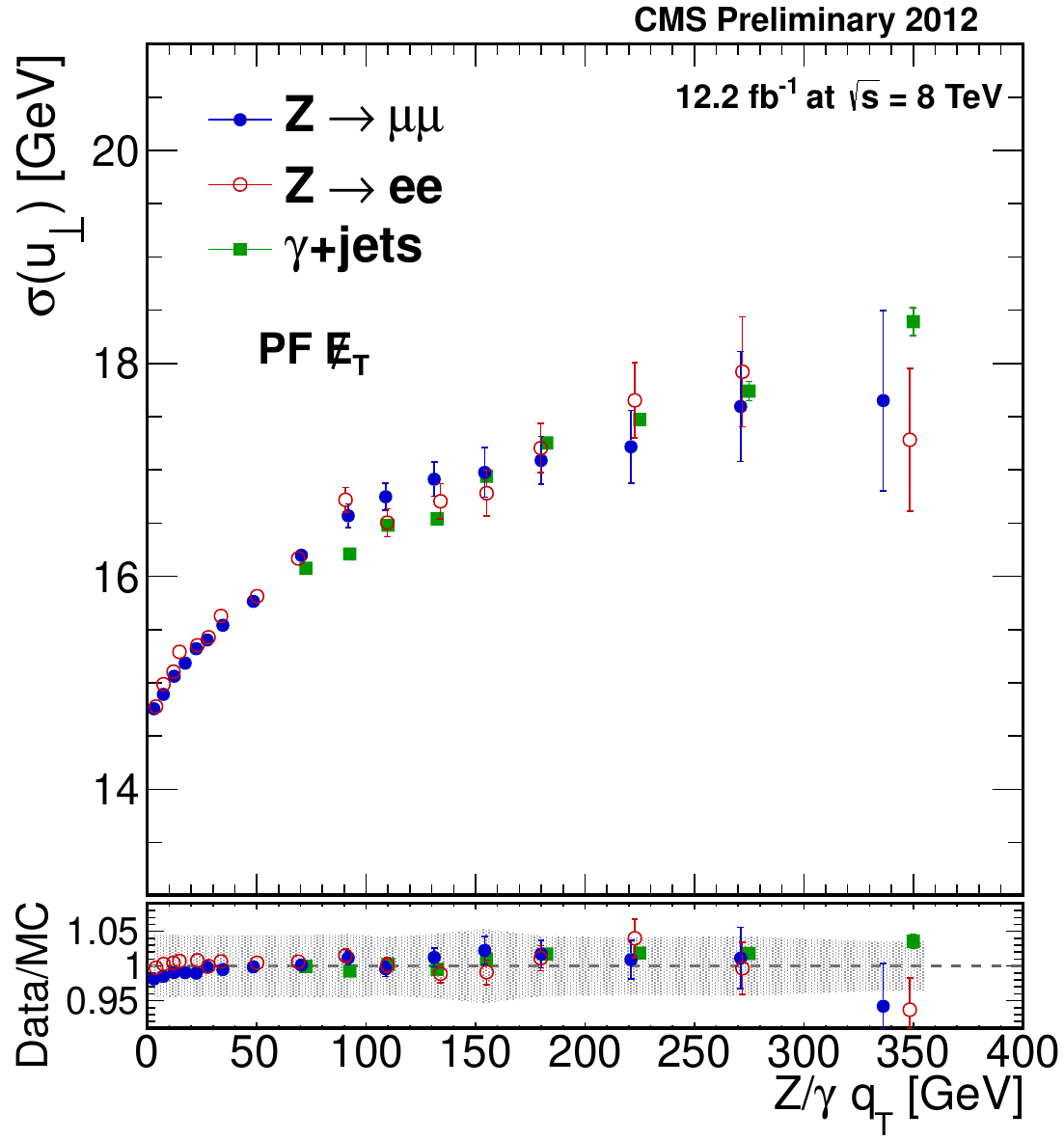}
\caption{Resolution of parallel recoil component (left) and perpendicular recoil component (right) for PF $\vec{E\!\!\!/{_T}}$ vs $q_{T}$ for $Z\rightarrow\mu^{+}\mu^{-}$, $Z\rightarrow e^{+}e^{-}$  and $\gamma$ events
 in data (on top), with the ratio of data to simulation (shown below).
 }
\label{fig:fig_PFresoQt}
\end{figure}

The $\vec{E\!\!\!/{_T}}$ scale is defined as $-u_{\parallel}/q_{T}$. The $\vec{E\!\!\!/{_T}}$ scale as a function of $q_{T}$  is shown in Figure \ref{fig:fig_PFresponse} for the different types of events. The $\vec{E\!\!\!/{_T}}$ resolution is estimated from the spread of the  $u_{\parallel} + q_{T}$  and $u_{\perp}$ distributions and is shown in Figure \ref{fig:fig_PFresoQt}, as a function of $q_{T}$. For both $\vec{E\!\!\!/{_T}}$ scale and resolution, a good agreement between the different channels and between data and simulation is found.

\section{Pileup mitigated $\vec{E\!\!\!/{_T}}$}
\label{sec5NoPUMET}
\label{sec4EffectofPU}
          
\begin{figure}[htbp]
\centering
\includegraphics[scale=.2]{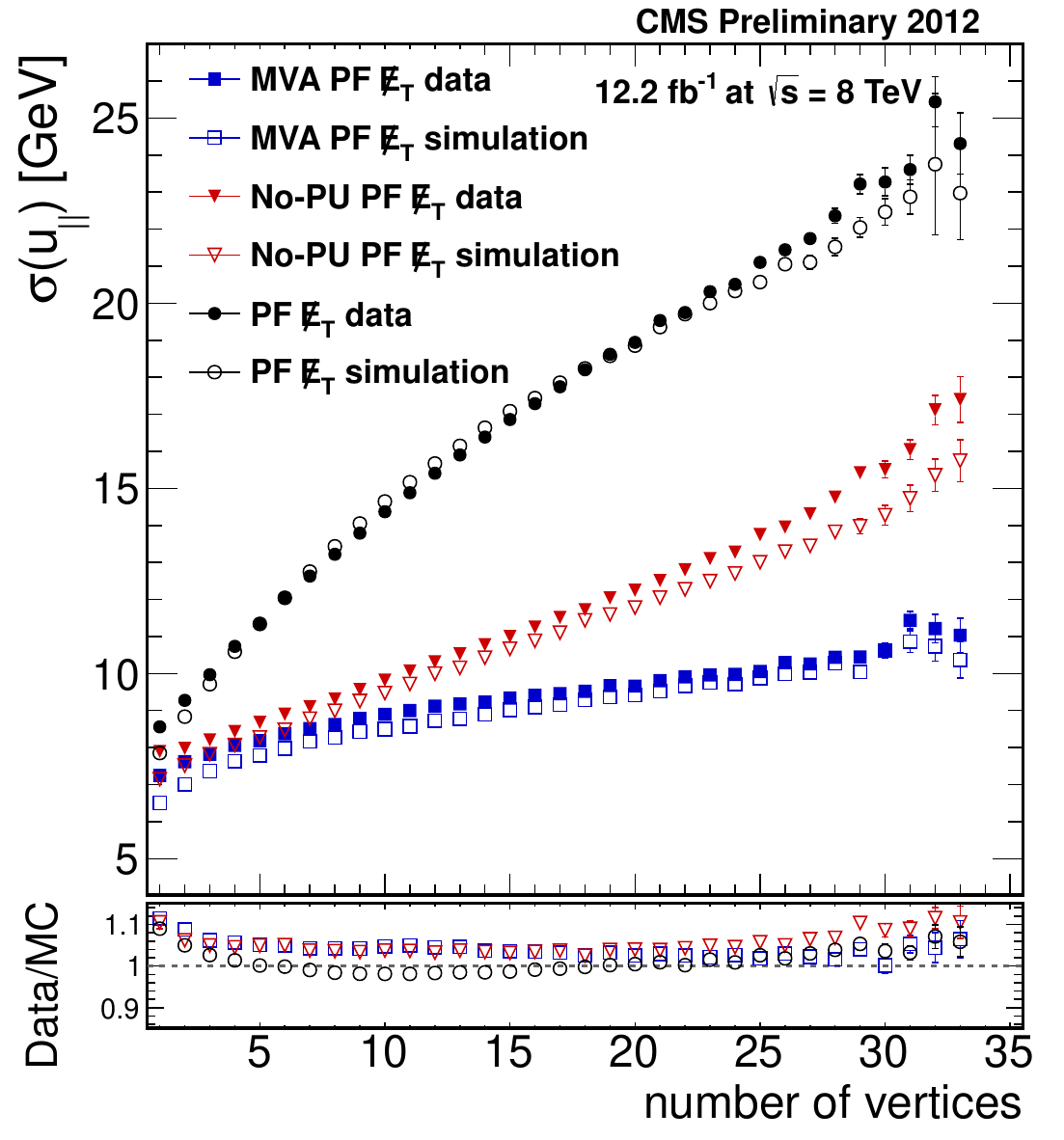}
\includegraphics[scale=.2]{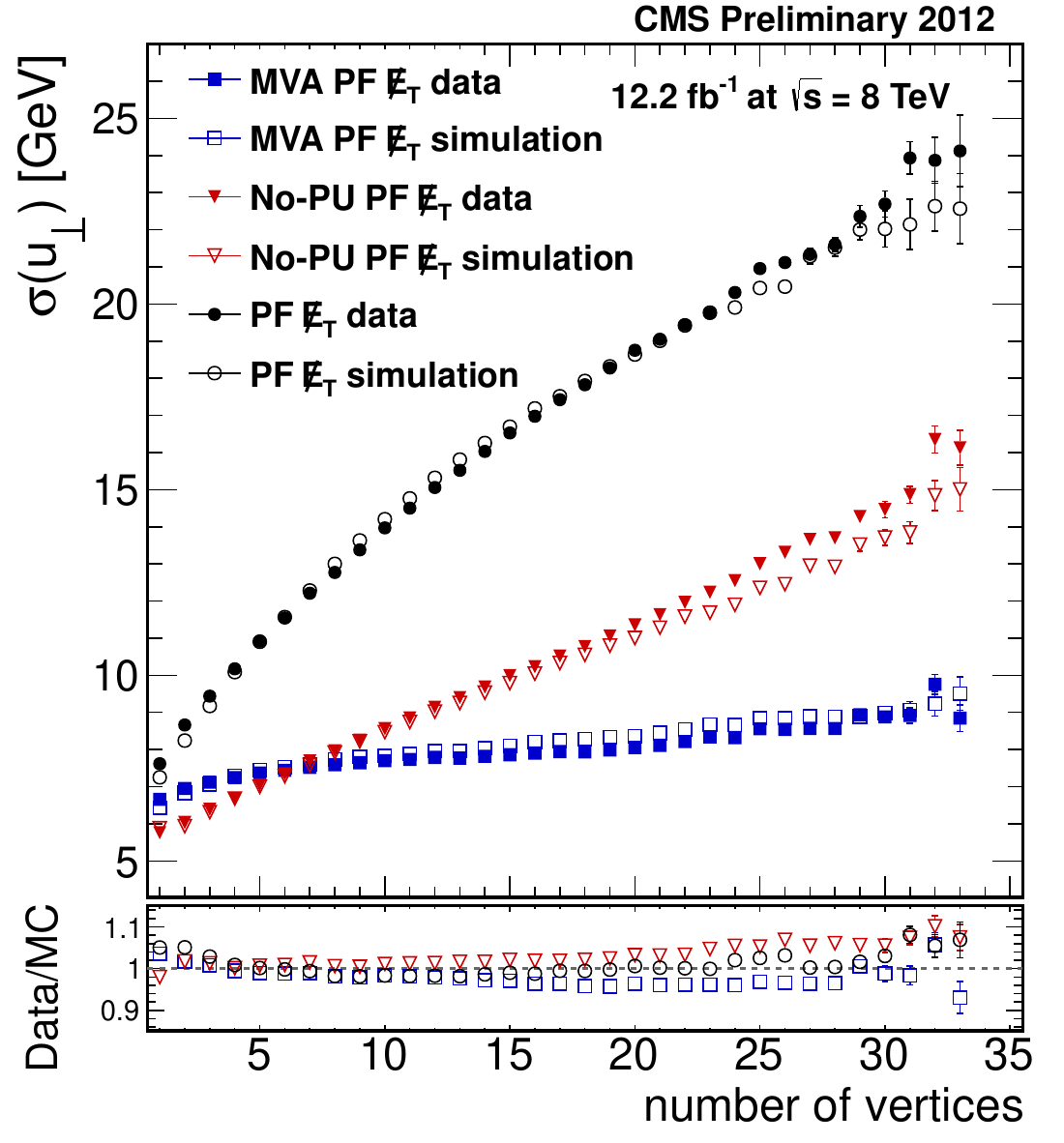}
\caption{Parallel (left) and perpendicular (right) resolution as a function of the number of reconstructed vertices for PF $\vec{E\!\!\!/{_T}}$, No-PU PF $\vec{E\!\!\!/{_T}}$ and MVA PF $\vec{E\!\!\!/{_T}}$ for $Z\rightarrow\mu^{+}\mu^{-}$ events.
}
\label{fig:fig_Zuu_uperpCombine_ResoNoPUMV}
\end{figure}

During the LHC runs in 2012, the average number of pileup interactions during proton-proton collisions was 21. Pileup mostly impacts $\vec{E\!\!\!/{_T}}$ resolution. In Figure \ref{fig:fig_Zuu_uperpCombine_ResoNoPUMV}, the $\vec{E\!\!\!/{_T}}$ resolution is shown as a function of pileup level estimated by the number of reconstructed vertices. For the PF $\vec{E\!\!\!/{_T}}$, each pileup interaction induces a deterioration in resolution of 3.3-3.7 GeV in quadrature.
\\
Two algorithms, No-PU PF $\vec{E\!\!\!/{_T}}$ and MVA PF $\vec{E\!\!\!/{_T}}$, have been developed to reduce the effect of pileup interactions on $\vec{E\!\!\!/{_T}}$. These algorithms are based on identifying and separating the PF particles that come from the primary hard-scattering proton-proton interaction (HS particles) and those that originate from pileup interactions (PU particles). Both algorithms make use of an MVA based pileup jet identification discriminator \cite{bib5CMSPUjetID}, that is able to separate jets from HS and jets from PU.
\\
No-PU PF $\vec{E\!\!\!/{_T}}$ is obtained by separating particles into those from HS (consisting of leptons, jets from HS and charged hadrons from HS) and those from PU (charged hadrons from PU, PU jets, neutrals not in jets) and then assigning different weights for their contribution to $\vec{E\!\!\!/{_T}}$.
\\
MVA PF $\vec{E\!\!\!/{_T}}$ provides an improved measurement of the $\vec{E\!\!\!/{_T}}$ by correcting the hadronic recoil $\vec{u_{T}}$ that is calculated from PF particles. Five different types of $\vec{E\!\!\!/{_T}}$ are evaluated taking different sets of PF particles and jets based on whether they originate from the HS or PU and these are used to correct the hadronic recoil and then the $\vec{E\!\!\!/{_T}}$.
\\
The $\vec{E\!\!\!/{_T}}$ resolution of No-PU and MVA PF $\vec{E\!\!\!/{_T}}$ functions of the level of pileup are also shown in Figure \ref{fig:fig_Zuu_uperpCombine_ResoNoPUMV}. The No-PU PF $\vec{E\!\!\!/{_T}}$ and particularly the MVA PF $\vec{E\!\!\!/{_T}}$ show a significant improvement in the pileup dependency of $\vec{E\!\!\!/{_T}}$ resolution compared to the PF $\vec{E\!\!\!/{_T}}$.
 
\section{Conclusion}
\label{sec6Conclusion}

The performance of $\vec{E\!\!\!/{_T}}$ reconstruction algorithms has been studied using data corresponding to an integrated luminosity of 12.2 $\mathrm{fb^{-1}}$ collected in 8 TeV pp collisions with the CMS detector at the LHC. 
The response and resolution of PF $\vec{E\!\!\!/{_T}}$ and the effect of pileup interactions on PF $\vec{E\!\!\!/{_T}}$ have been measured.
The distributions in data and simulation agree well.
 Studies of two new $\vec{E\!\!\!/{_T}}$ algorithms, No-PU PF $\vec{E\!\!\!/{_T}}$ and MVA PF $\vec{E\!\!\!/{_T}}$, show a significantly reduced dependence of the $\vec{E\!\!\!/{_T}}$ resolution on pileup interactions.




\nocite{*}
\bibliographystyle{elsarticle-num}
\bibliography{martin}



\end{document}